\documentclass[useAMS,usenatbib]{mn2e}
\usepackage{lscape}
\usepackage{graphicx,times}
\usepackage{float}
\newcommand{\lesssim}{\mbox{{\raisebox{-0.4ex}{$\stackrel{<}{{\scriptstyle\sim}}$}}}}
\newcommand{\gtrsim}{\mbox{{\raisebox{-0.4ex}{$\stackrel{>}{{\scriptstyle\sim}}$}}}}

\newcommand{\bnt}{\left({B\over 1\,{\rm nT}}\right)}
\newcommand{\bequipnt}{\left({B_{\rm equip}\over 1\,{\rm nT}}\right)}
\newcommand{\gamthous}{\left({\gamma_{\rm min}\over 10^3}\right)}
\newcommand{\nzero}{\left({N_0\over 4.3\times 10^{-2}\,{\rm m^{-3}}}\right)}
\newcommand{\numin}{\left({\nu_{\rm min}\over 8\,{\rm GHz}}\right)}

\title[X-ray evidence of relic jet activity in Cygnus A]{Multiwavelength study of Cygnus A II. X-ray
  inverse-Compton emission from a relic counterjet and implications
  for jet duty-cycles} 
\author[K.C.\ Steenbrugge,  K.M.\ Blundell and P.\ Duffy]{Katrien
  C. Steenbrugge$^{1}$\thanks{E-mail:kcs@astro.ox.ac.uk}, Katherine
  M. Blundell$^{2}$ and Peter Duffy$^{3}$ \\
$^{1}$St John's College Research Centre, University of Oxford, St
  John's College, Oxford, OX1 3JP, UK\\ 
$^{2}$University of Oxford, Department of Physics, Keble Road, Oxford,
  OX1 3RH, UK\\
$^3$ UCD School of Mathematical Sciences, UCD, Dublin 4, Ireland}

\begin{document}

\date{Accepted . Received }

\pagerange{\pageref{firstpage}--\pageref{lastpage}} \pubyear{2002}

\maketitle

\label{firstpage}

\begin{abstract}
The duty-cycle of powerful radio galaxies and quasars such as the
prototype Cygnus\,A is poorly understood.  X-ray observations of
inverse-Compton scattered Cosmic Microwave Background (ICCMB) photons
probe lower Lorentz-factor particles than radio observations of
synchrotron emission and thus potentially reveal a more aged
population. Comparative studies of the nearer and further lobes,
separated by many 10s of kpc and thus 10s of thousands of years in
light-travel time, yield additional temporal resolution in studies of
the lifecycles of such objects.  We have co-added all archival {\it
Chandra} {\rm ACIS-I} data and present a deep 200\,ks image of
Cygnus\,A.  This deep image reveals the presence of X-ray emission
from a counterjet i.e.\ a jet receding from Earth and related to a
previous episode of jet activity. The outer part of this counterjet
does not overlay the current counterjet detected in radio emission,
excluding the possibility that we detect the current counterjet in
X-rays. The non-thermal X-ray emission has a power-law photon index is
1.7, and we interpret this emission as ICCMB radiation.  There is an
absence of any discernible X-ray emission associated with a jet
flowing towards Earth.  We conclude that: (1) The emission from a
relic jet, indicates a previous episode of jet activity, that took
place earlier than the current jet activity appearing as synchrotron
radio emission. (2) The presence of X-ray emission from a relic
counterjet of Cygnus\,A and the absence of X-ray emission associated
with any relic approaching jet constrains the timescale between
successive episodes of jet activity to $\sim 10^6$ years.  (3)
Transverse expansion of the jet causes expansion losses which shifts
the energy distribution to lower energies. Particles with initially
high Lorentz factors, that originally gave detectable synchrotron
radiation, attain Lorentz factors $\sim 10^3$ and inverse-Compton
scatter CMB photons, to give X-ray emission.  (4)Assuming the
electrons cooled due to adiabatic expansion, the required magnetic
field strength is substantially smaller than the equipartition
magnetic field strength. (5) A high minimum Lorentz factor for the
distribution of relativistic particles in the current jet, of a few
$10^3$, is ejected from the central nucleus of this active
galaxy.
\end{abstract}


\section{Introduction}

The Cygnus\,A cluster and galaxy (3C\,405) are one of the brightest
sources in the X-ray sky and have therefore been studied with every
major X-ray satellite. In this paper we take advantage of the high
spatial resolution of the {\it Chandra} satellite to study the
linear counterjet-like feature this reveals.

The spatial resolution of the ACIS camera onboard {\it Chandra} allows
us to spatially resolve the jet, lobes and hotspots from the central
AGN. Therefore one can compare the radio and X-ray properties of a
particular galaxy with fine spatial resolution. Cygnus\,A was observed
11 times with {\it Chandra}, 10 of which have an exposure time of more
than 5 ks, and are used in this paper.  Some of the {\it Chandra}
datasets we use have been analysed and published by \cite{wilson00},
\cite{young02}, \cite{smith02}, \cite{balucinska-church05},
\cite{croston05}, \cite{evans06} and \cite{wilson06}.

One question addressed in this paper relates to the duty cycle of the
jet-activity in Cygnus\,A.  That is, of the time for which the super
massive black hole central engine in Cygnus\,A is accreting, for what
fraction of that time is matter expelled in the form of relativistic
jets?  In a recent paper, Nipoti, Blundell \& Binney (2005) contended
that radio-loudness (manifested by jet activity) in active galaxies
such as quasars was analogous to the intermittency of the jet ejection
in microquasars, albeit on longer timescales.

Detection of X-ray photons that arise from inverse Compton
up-scattered Cosmic Microwave Background (ICCMB) photons mandates the
presence of relativistic particles with Lorentz factors of order
$10^3$ \citep{harris79}.  Such particles are likely to have lower
Lorentz factors than ambient synchrotron-emitting particles radiating
at the typically-observed radio wavelengths, assuming the magnetic
field strengths in the lobes of radio galaxies are nT in size or
lower.  Thus, co-spatial X-rays can reveal information about the
lower-energy population of a distribution of relativistic particles
than synchrotron radio emission from the same plasma. Furthermore, they
can signal the presence of relic (that is previously, but no longer
detectable synchrotron emitting) plasma (e.g. \cite{erlund06}, and
\cite{blundell06}).  Examination of the brightness distribution of
ICCMB --- relative to synchrotron --- gives an extra step in temporal
resolution in these objects which evolve slowly relative to human
timescales.

For a redshift of 0.05607 \citep{owen97} the physical size, i.e.
the distance between the outer hotspots not correcting for possible
line of sight angle, of Cygnus\,A is 130\,kpc; assuming a cosmology
with $H_0$ = 73 km s$^{-1}$ Mpc$^{-1}$ and $\Omega_{\rm M} = 0.3$ and
$\Omega_\Lambda$ = 0.7. Therefore, the light-travel time
between opposite lobes exceeds $\cos \theta \times 4 \times 10^5$
years, where $\theta$ is the angle between the axis of the radio
source and our line-of-sight.  Since the light we observe from
opposite lobes is received at the same telescope time, this means
that an observer on Earth sees the nearer lobe at a more recent
epoch than the further lobe, which is seen at an earlier time in the
radio galaxy's history.  Properly accounting for light-travel time
effects is important in the interpretation of side-to-side
asymmetries in the lobes and jets of quasars and microquasars
\citep[e.g.\ ][]{blundell94,miller-jones04}.

An important corollary of the different
epoch at which we observe different sides of the source is that ---
relative to the near lobe --- we are looking back in time, and
attaining an extra, different, step in temporal resolution in these
slowly evolving objects.

\section{Observations and data reduction\label{sect:obs}}

There were 3 observing campaigns on Cygnus\,A with {\it Chandra}
resulting in 10 observations with an exposure time of at least 5
ks. The details of the observations are listed in
Table~\ref{tab:obs}. All observations used the ACIS (Advanced CCD
Imaging Spectrometer) instrument. The first 2 observations utilized
the ACIS-S, therefore the image of Cygnus\,A fell on a
back-illuminated chip, which has a higher effective area for lower
energies. In the second observation a subarray of the CCDs was
illuminated, thus allowing for a read-out time of only 0.4 s. This
set-up minimizes the pile-up in the core of the AGN. In the other 9
observations the core is seriously piled-up. The last 8 observations
are with the ACIS-I configuration and in the VFAINT mode, which gives
a reduced background after processing. All the data were obtained from
the {\it Chandra} public archive and reduced (including the thread
to obtain the reduced background) with the standard threads 
  (a collection of commands) in CIAO
3.3 (http://asc.harvard.edu/ciao/threads/), which included the updated calibration database CALDB 3.2.2. 
  The filtering minimally reduced the exposure times,
  Table~\ref{tab:obs} lists the filtered exposure times. The
background region was chosen from a low count rate region on the CCD
array containing the image of Cygnus\,A; however, the position of the
background region in observation 1 and 2 is different from that of the
remaining observations, due to the different instrumental set-up. A
circle with radius 14.76$^{\prime\prime}$ centered on
19$^h$59$^m$41$^s$.335 in Right Ascension (J2000) and
+40$^\circ$40$^{\prime}$51$^{\prime\prime}$.03 in Declination (J2000)
was used for all but observation 2. For the 2$^{\rm nd}$ observation,
which has only a strip of the CCDs exposed, a circle centered on
19$^h$59$^m$46$^s$.172 in Right Ascension and
+40$^\circ$39$^{\prime}$46$^{\prime\prime}$.06 in Declination with a
radius of 12.3$^{\prime\prime}$ was used.

We aligned the AGN core detected in the 2 $-$ 10 keV band of the first
X-ray observation with the fitted coordinates for the 5-GHz core, from
observations made by \citet{carilli91}.  The 2 $-$ 10 keV band was
chosen to avoid the extended soft X-ray emission detected by
\cite{young02}. Matching the nucleus accurately led to the
superposition of the radio and X-ray detected hotspots. This gives us
confidence that the radio and X-ray core of the AGN indeed coincide
within the resolution of {\it Chandra}. Using the coordinates
determined for the first observation we used the reproject$\_$aspect
thread to re-align the other 9 observations. This method resulted in
an alignment of the core to within 0.5$^{\prime\prime}$ for 7 of the
observations, but failed for observations 2 and 6. For observation 2
we did not detect the necessary number of point sources on the strip
of the CCDs exposed. A similar problem occurs for observation 6,
however, in this case it is due to the short exposure time and not the
instrumental set-up. For observation 2 we used AIPS
(http://www.aips.nrao.edu/) to fit a Gaussian to the core and then
matched the position of the peak in emission to that of the radio
determined core. This resulted again in an alignment to better than
0.5$^{\prime\prime}$. This method however failed for observation 6;
therefore we readjusted the coordinates until alignment was achieved
between the pixel with maximum counts in observation 6 and that in
observation 1.

\begin{table}
\begin{center}
\caption{The list of observations of Cygnus\,A used in this paper,
  i.e. all {\it Chandra} observations with an filtered exposure
  time longer than 5 ks. Listed are the date of the observation, the
  exposure time as well as the instruments used: ACIS-S or ACIS-I, the mode of
  observation  and ObsId number. See main text for more details.} 
\label{tab:obs}
\begin{tabular}{|l|l|c|l|l|l|}\hline
    &   date     & exposure (ks) & instrument  & mode        & ObsID \\\hline
1   & 2000 05 21 & 34.72         & ACIS-S      & FAINT       &  360  \\
2   & 2000 05 26 & 10.17         & ACIS-S      & 0.4 s frame &  1707 \\
3   & 2005 02 15 & 25.80         & ACIS-I      & VFAINT      &  6225 \\
4   & 2005 02 16 & 51.09         & ACIS-I      & VFAINT      &  5831 \\
5   & 2005 02 19 & 25.44         & ACIS-I      & VFAINT      &  6226 \\
6   & 2005 02 21 & 6.96          & ACIS-I      & VFAINT      &  6250 \\
7   & 2005 02 22 & 23.48         & ACIS-I      & VFAINT      &  5830 \\
8   & 2005 02 23 & 23.05         & ACIS-I      & VFAINT      &  6229 \\
9   & 2005 02 25 & 16.04         & ACIS-I      & VFAINT      &  6228 \\
10  & 2005 09 07 & 29.65         & ACIS-I      & VFAINT      &  6252\\\hline   
\end{tabular}
\end{center}
\end{table}

\begin{figure}
\begin{center}
  \resizebox{\hsize}{!}{\includegraphics[angle=0]{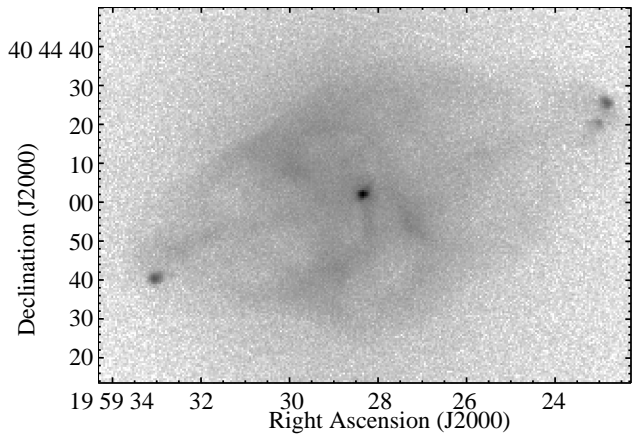}}
  \caption{The 0.2--10\,keV ACIS-I image of Cygnus\,A. This image was
  obtained by adding the last 8 observations listed in Table\,1, after
  reprojecting the files so that the coordinates matched the 5-GHz
  radio coordinates. \label{fig:merge}} 
\end{center}
\end{figure}

Once all 10 X-ray observations were aligned with the 5~GHz radio
image, we added the 0.2--10\,keV images producing one image for the
ACIS-S exposures and one for the ACIS-I exposures using the
merge$\_$all command in CIAO. We used the first and fifth observations
to provide the reference coordinates for the ACIS-S and ACIS-I images
respectively.  The resulting ACIS-I image, which has the better
statistics of the 2 images, is shown with different transfer functions
in Figs~\ref{fig:merge} and \ref{fig:x-ray_cjet}. For extraction of
the spectra from different regions we used the specextract command in
CIAO. Considering the difference in detector and/or observation mode
as well as the time span between the different observations we
extracted all the spectra for the different regions for each
observation, using the specific badpixel file for that
observation. The quoted errors on the X-ray luminosity and photon
index is for $\Delta\chi^2$ = 2, the RMS of the $\Delta\chi^2$
distribution (or a confidence level of 84.3 \% for 1 free parameter)
\citep{kaastra04}. We used the SPEX \citep{kaastra02b} package for
fitting the spectra.

\section{Results\label{sect:results}}
\subsection{X-ray counterjet\label{sect:x_jet}}

\begin{figure}
\begin{center}
 \resizebox{\hsize}{!}{\includegraphics[angle=0]{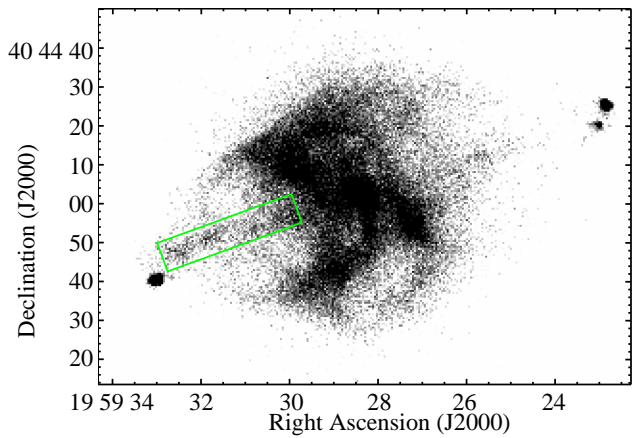}}
\caption{The 0.2$-$10\,keV ACIS-I image, with a different transfer
  function from that of Fig.\,1, clearly showing the linear
  counterjet-like feature delineated by the green box and the
  non-detection of anything corresponding to this on the jet
  side.
  \label{fig:x-ray_cjet}} 
\end{center}
\end{figure}

Interestingly, a linear counterjet-like feature is easily detected in
the 0.2$-$10 keV image (see Fig.~\ref{fig:x-ray_cjet}), as first
reported by \cite{steenbrugge07}. This seems to be most prominent
further away from the nucleus and bends at RA=19$^h$59$^m$31$^s$ and
Dec=+40$^\circ$43$^{\prime}$54$^{\prime\prime}$. This bend occurs just
inside the two weaker jet knots observed in the 15~GHz radio image (E4
and E5 in Fig.\,6 of Steenbrugge \& Blundell, accepted companion
paper) and between which the counterjet starts to bend over a large
angle. We use a box (see Fig.~\ref{fig:x-ray_cjet}) with centre
19:59:31.358 in RA and +40:43:52.54 in Declination, a length of
36.71$^{\prime\prime}$ and width of 7.79$^{\prime\prime}$ and an angle
counter clock-wise of  20$^{\circ}$ from east to fit the
counterjet feature. This is the only significantly detected
feature that lies partly within the 5~GHz lobes. Inevitably, the box
will contain some emission from the background thermal gas originating
from the cluster. The width of the counterjet is
$\sim$5$^{\prime\prime}$, and is thus resolved in our {\it Chandra}
image. The width of the brightest knots observed in the 15~GHz image,
which have the largest width of the knots in any of the three radio
bands we use, are $\sim$2$^{\prime\prime}$.9 (Steenbrugge \& Blundell,
accepted companion paper). Thus the X-ray counterjet feature is wider
than any radio jet knots in Cygnus~A.

\subsection{Fitting the X-ray counterjet spectrum}
 The X-ray counterjet is observed against a bright and variable
  local background, due to thermal gas in close vicinity of the
  galaxy. Therefore, we decided to fit this local background as an
  extra component in our fit to the spectrum of the counterjet, rather
  than subtracting an unknown local background.
The X-ray spectrum of the counterjet (as indicated in
Fig.~\ref{fig:x-ray_cjet}), is well fitted (reduced $\chi^2$ = 1.1 for 1582
degrees of freedom) by a power-law with Galactic absorption of $3.5
\times 10^{25} {\rm m}^{-2}$ \citep{dickey90}. We rebinned the
spectrum by a factor of 3, and fitted the spectrum between 0.5$-$7
keV. The 0.1$-$10 keV luminosity is (1.4 $\pm$0.2)$\times$10$^{36}$~W
or (7.0 $\pm$ 0.12)$\times$10$^{35}$ W in the 2$-$10 keV band. The
normalization ( normalized at 1 keV) is 1.80$\times$10$^{51}$
photons s$^{-1}$ keV$^{-1}$, and the photon index is 1.70 $\pm$
0.02. There is likely to be contamination of the spectrum by the
surrounding hot cluster gas, but the data are too poor to constrain
   the temperature and normalization of this component, which
  we fitted as an extra component, using the 
  {\it cie} model in SPEX. A good indication of the luminosity of this
contamination can be obtained from the derived luminosity of the
thermal component, fitted together with a power-law, for the jet. The
thermal luminosity is 2.1$\times$10$^{35}$ W in the 2$-$10 keV
  range. Fitting the counterjet feature with a 
thermal model gives a poorer fit, namely a reduced $\chi^2$ = 1.2 (for the same
degrees of freedom, for a temperature of 6.3 keV and an emission
measure of $5.59 \pm 0.13 \times 10^{71} {\rm m}^{-3}$).
\cite{wilson06} studied the brighter features, although not the counterjet,
in Cygnus A and convincingly showed that the higher temperature gas is
at the outer edge, i.e. the contact discontinuity. The gas more
centrally located is quite a bit cooler, having temperatures ranging
  between 3.80 and 4.28 
keV and can be explained as being due to the jet break-out phase as
modelled by \cite{sutherland07a} and \cite{sutherland07b}. The
location of the counterjet-like feature that we observe in the X-rays
  is not consistent with it either 
belonging to the gas heated by the jet break-out or the contact
discontinuity, a reason why a thermal explanation is less likely.


\begin{figure}
\begin{center}
 \resizebox{\hsize}{!}{\includegraphics[angle=0]{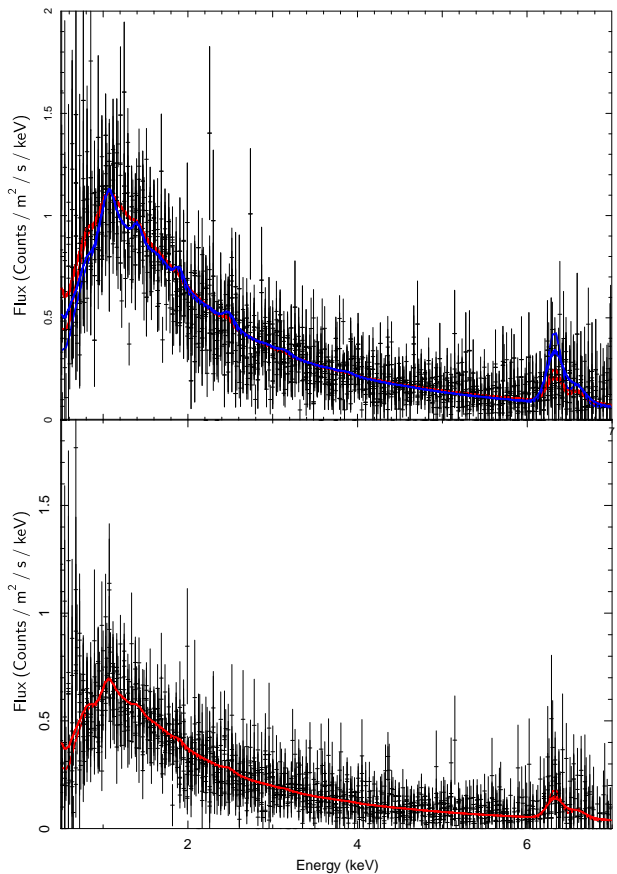}}
\caption{ Flux spectra for the energy interval 0.5 $-$ 7~keV from the
    10 different observations.  {\em Upper panel:} Spectrum from the
    counterjet feature within the eastern lobe and the best fit
    power-law plus thermal model in red. The blue curve
    indicates the best fit thermal model. Note that the 6.4
    Fe~K$\alpha$ line is severly overpredicted in this fit.  The
    slight excess around 6.4 keV is due to the thermal emission from
    the cluster surrounding Cygnus~A. The small difference between the
    fits at shorter wavelengths is a direct result of the different
    calibrations for the different set-ups of the instruments used.
    {\em Lower panel:} Spectrum from a corresponding region on the
    western lobe where the current jet is observed.  The fit to the
    data shown is a power-law plus thermal model with a spectral slope
    of 1.75 and temperature of 4 keV. The small differences between
    the fits at shorter wavelengths and at 6.4 keV is a direct result
    of the different calibrations and resolution for the different
    set-ups of the instruments used.
  \label{fig:bothjets}} 
\end{center}
\end{figure}


The spectrum of the counterjet-like feature is in contrast to the mainly
thermal spectrum of the other bright, central and curved features
clearly seen in Fig.~\ref{fig:x-ray_cjet}. The X-ray counterjet obeys
the \citet{bridle84} criteria for jets: its length is more than four
times larger than its width; it is separable at high spatial
resolution from the surrounding features; and it is aligned with the
compact core. Therefore, on the basis of these and its power-law
spectrum, we conclude that the long linear feature identified on the
east of the source is indeed related to jet activity and not part of
the surrounding environment. Furthermore, the counterjet lies on
  the same line that connects the brighter western hotspot (i.e. the
  brighter hotspot on the jet side) and the nucleus.

\par The inner part of the X-ray counterjet does overlie the
inner part of the counterjet detected in the radio images, however it
does not make the 27$^{\circ}$31$^{\prime}$ bend observed in the
15-GHz image (Steenbrugge \& Blundell, accepted companion paper);
rather it extends along its original direction until just north of the
bright eastern hotspot. There is a clear gap in emission between the
end of the X-ray detected counterjet and the hotspot. We conclude that
the X-ray detected counterjet is a relic for the following three
reasons: (i) it is extended transversely compared to the radio
counterjet and radio jet (as explored in Sect.~\ref{sect:cooling}), (ii)
it does not overlay the outer radio 
counterjet, (iii) there is a gap in emission between the observed
counterjet and the hotspots.

\par The inner part of the relic counterjet overlaps with the inner
part of the current 
counterjet, and therefore we cannot constrain in this region the X-ray
luminosity potentially coming from the inner current jet. However, for
the outer X-ray counterjet there is a lack of associated 15-GHz radio
emission (see fig.~7, Steenbrugge \& Blundell, accepted companion
paper). This indicates a lack of high Lorentz-factor particles in this
relic counterjet compared with those observed in active jets. 

\subsection{Limits on the density of thermal gas within the lobes}

We now consider the possibility of whether the counterjet feature
could be explained via thermal emission.  We find in
Section\,\ref{sect:x_jet} that the emission measure for a thermal
plasma is $5.59 \pm 0.13 \times 10^{71} {\rm m}^{-3}$.  From
this emission measure we derive an electron density of $3.1 \times
10^5 {\rm m}^{-3}$, assuming $n_{\rm e} = 2.1 n_{\rm H}$, using $Y =
n_{\rm e}n_{\rm H}V$ and a volume of $1.22 \times 10^{61}\,{\rm
m}^{3}$.  The upper limit to the thermal electron density in the
western lobe detected in 5~GHz of Cygnus\,A is derived by
\citet{dreher87} to be $4 
\times 10^2\,{\rm m}^{-3}$ for an isotropic random magnetic field 
from the lack of depolarization observed at eight points in the lobes;
this could be higher by as much as two orders of magnitude if the
magnetic field has many reversals.  Even this extreme upper limit
still falls short of the necessary electron density, for the emission
to be thermal, by one order of magnitude.  This rules out that the
feature is due to thermal emission in the lobe. Considering that
Cygnus~A cluster is a relaxed cluster, showing no sharp, delineated
features outside the volume around the Cygnus~A galaxy bounded by the
lobes and hotspots, we conclude that this feature is unlikely to be
thermal,  either from the cluster or the gas surrounding the galaxy.

\subsection{Limits on X-ray emission from an approaching jet}

Tracing the inner part of the jet or counterjet in the X-rays is not
straightforward, due to the hot thermal gas surrounding the nucleus
and thereby possibly hiding the emission from any inner jet. There is
excess X-ray emission centred on Right Ascension=19$^h$59$^m$25$^s$.680 and
Declination=+40$^{\circ}$44$^{\prime}$12$^{\prime\prime}$.36, in the
western lobe. This feature lies to the north of a jet knot detected at
5\,GHz, and therefore is unlikely to be current jet emission.  It is
located just inside the inner edge of the lobe; a possible explanation
for this feature is that X-ray thermal gas is colliding with the
backflowing lobe gas. However, we cannot exclude it being a very faint
trace of emission from a relic approaching jet. We conclude that in
Cygnus\,A the current jet is not detected in the 0.2--10\,keV X-ray
image. We inspected the 6$-$10\,keV image for any evidence of any
relic material arising from an approaching 
jet. There is a slight excess in detected counts just north of the
radio jet, overlapping with the lobe. The excess is most pronounced
compared with the area just south of the radio jet.  We note that
there is clear excess of radio emission to the north of the western
lobe (see fig\,1, Steenbrugge \& Blundell, accepted companion paper).

The 3 $\sigma$ upper limit to the 0.1--10\,keV luminosity for
any approaching jet (see Fig.~\ref{fig:bothjets}), assuming the
spectrum of a relic or current jet is a power-law (corrected for
Galactic absorption) is 1.6$\times$10$^{35}$ W, which has a best-fit
spectral slope of 1.75 $\pm$ 0.02.  The box for the jet was centered on
  RA: 19$^h$59$^m$25$^s$.507 and Dec.:
  40$^{\circ}$44$^{\prime}$12$^{\prime\prime}$.76, with a 
length of 33.76$^{\prime\prime}$ and the same width as the counterjet
box, rotated by 20$^{\circ}$. The normalisation of the power-law was
6$\times$10$^{50}$ photons s$^{-1}$ keV$^{-1}$. A power-law-only fit
to the data in the western lobe gives a reduced $\chi^2$ of 1.24 (for
1488 degrees of 
freedom), while a thermal-only model yields a reduced $\chi^2$ of 1.2
(same d.o.f), while a fit including a thermal and power-law component
gives a reduced $\chi^2$ of 1.1 (for 1486 d.o.f). The upper limit to
the power-law luminosity is determined by subtracting the luminosity
in a box of the same size but offset from the radio detected jet. The
width of the box is the same as the width for the box for the detected
counterjet. The spectrum obtained from the box centered on the radio
jet does show a weak Fe K$\alpha$ emission line, indicating that at
least part of the emission is thermal. This is further indication that
the flux detected is from the surrounding lobe or cluster and not an
approaching jet, or even a relic jet.

Note that this upper limit to the jet luminosity is rather large, due
to the presence of hot gas surrounding the Cygnus\,A galaxy from the
intracluster gas. This upper limit is only a quarter of the luminosity
detected for the counterjet.

\subsection{Other bright features in the X-ray image}

Figs.~\ref{fig:merge} and \ref{fig:x-ray_cjet} clearly show that there
are several bright extended regions (besides the counterjet, the
hotspots and the nucleus) which are well fitted with a thermal
component. This is completely consistent with the results obtained by
\cite{wilson06}. A thorough analysis of the bright extended
thermal emission is given by \cite{wilson06} and \cite{smith02}.
Therefore we can exclude ICCMB as the emission mechanism for these
bright regions.  In a future paper, we will discuss the thermal
features of the X-ray emission associated with Cygnus\,A in
detail. The structure of these bright features is
consistent with having either originated in the shock break-out of the
jet as modelled by \cite{sutherland07a} and \cite{sutherland07b} or
being the contact discontinuity as studied by \cite{wilson06}.

\subsection{Comparison with other X-ray detected jets and lobes}
\label{sect:other}

We do not detect in the X-ray image of Cygnus\,A any of the jet knots
observed in the different radio bands, either in the jet or the
counterjet.  Moreover, the X-ray counterjet-like feature has a
slowly varying brightness distribution along its length, in contrast
with the well-studied knotty X-ray emission from jets generally
attributed to synchotron radiation, and perhaps more in keeping with
what might be expected for ICCMB.

Pictor\,A is the only other radio galaxy with a possible detected
X-ray counterjet. \cite{hardcastle05} explain this weak counterjet
emission as synchrotron emission, because the X-ray flux ratio between
the jet and counterjet is about 6; and the X-ray spectrum of the
counterjet is steep. In Pictor\,A the X-ray jet is clearly more
luminous than the counterjet, contrary to the case in Cygnus\,A.  In
Pictor\,A there is no indication of a radio counterjet in VLBI
observations \citep{tingay00}. For Cygnus A \cite{krichbaum98}
do detect the VLBI counterjet.

The relative smoothness of the X-ray emission coming from the
counterjet is is very different from, for example, 3C\,303 (an unusual
double radio source, \cite{Kataoka03}), 4C\,19.44 (a Seyfert 1 galaxy
according to Simbad, \cite{sambruna02}), Cen\,A (prototypical FR I
radio galaxy, \cite{kraft02}), and M\,87 (an FRI, \cite{marshall02}),
where in the X-rays individual knots are clearly seen and the
brightness of the jet is rather variable but much less
luminous. Furthermore, in these jets, the knots closest to the nucleus
are in general the brightest in X-rays and the emission is generally
attributed to the synchrotron mechanism.  In constrast, in Cygnus\,A
the brightness is not much weaker at the furthest end of the
counterjet. A further discussion of these important differences is
deferred to a future paper.

For a number of radio galaxies, the counterlobe is brighter in
X-rays than the lobe (\cite{brunetti01}; \cite{brunetti02};
\cite{sambruna02}; \cite{bondi04}). The asymmetric brightness is
explained as inverse-Compton (IC) emission, due to anisotropic
scattering of infrared and optical photons from the nucleus
\citep{brunetti97}. In all these cases the brightness of the X-ray lobe
falls off steeply with increasing distance from the
nucleus. Furthermore, the luminosity asymmetry is a strong function of the
inclination angle, and the counterjet needs to be inclined to the line
of sight by a rather large angle to see this effect clearly.  

\par According to \cite{brunetti97}, the IC emission from the
counterlobe in Cygnus\,A should be small compared to the thermal
component, and that is indeed consistent with our analysis of the
(counter)lobe spectra.

\section{Discussion}

\subsection{Mechanism for generation of ICCMB and implications for the
  low-energy turnover} 

The upper limit of ICCMB associated with the current episode of jet
activity (either from the jet or the counterjet) together
with the presence of ICCMB associated with previous jet activity,
implies an upper limit to the number density of $10^3$ Lorentz factor
particles.  This, together with the clearly evident current jets
emitting radio synchrotron, suggests that the current synchrotron jet
plasma has a low-energy turnover above $10^3$ in the distribution of
Lorentz factors of the plasma ejected from close to the black hole.

However, the absence of ICCMB from the current jet could potentially
be due in part to X-ray bright galaxy/cluster emission. The upper
limit to the X-ray brightness of the jet is just a quarter of that of
the counterjet luminosity, implying that the number density of
particles with Lorentz factors $\sim 10^3$ is less than a quarter in
the current jet compared with that in the observed relic
counterjet. The upper limit for the current jet is still about 450
times brighter than the jet in Cen~A, so a current jet of similar
luminosity will have gone undetected, as would the current
counterjet. However, the emission mechanism of the Cen A jet is
synchrotron emission, therefore sampling much higher Lorentz factor
particles.

\subsection{Light-travel time effects and cooling}

The observation of X-ray emission from the counterjet side, and the
absence of such emission from the approaching side of the source,
places general constraints on the rate at which the X-ray emission
must decline which, assuming symmetrical conditions, depends purely on
light-travel time arguments.  We consider approaching and receding
jets moving with speed $v=\beta_{\rm HS} c$ from the central source
and making an angle $\theta$ with the observer's line of sight. When
the approaching jet is observed to have extended out to a distance
$d$, so that $t_{\rm app}=d/v$ is the time since the jet started
advancing, we will observe emission from the furthest extremity of the
receding jet. The extremity, i.e.\ hotspot, of the receding jet is
observed at a younger age than that at the closest point to us of the
approaching jet, with the ratio of ages ($t_{\rm rec}$ and $t_{\rm
  app}$) given by

\begin{equation}
t_{\rm rec}=\left({1-\beta_{\rm HS}\cos\theta}\over{1+\beta_{\rm
    HS}\cos\theta}\right)t_{\rm app}. 
\end{equation}

Therefore, the difference in age between the observed jet extremities,
which must of course be imaged at the same ``telescope time'', is

\begin{eqnarray}
\displaystyle  \Delta t
=\frac{2\beta_{\rm HS}\cos\theta}{1 + \beta_{\rm HS}\cos\theta}t_{\rm
  app}. 
\end{eqnarray}

\begin{figure}
\begin{center}
 \resizebox{\hsize}{!}{\includegraphics[angle=0]{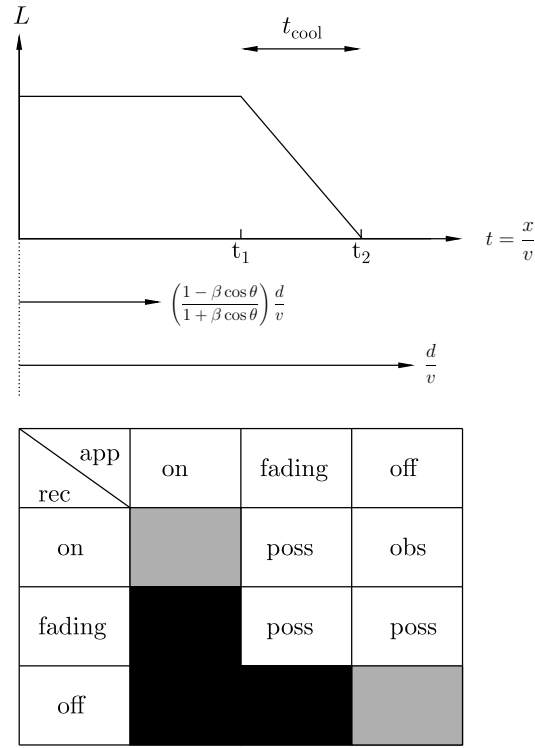}}
\caption{{\em Upper:} This is a schematic lightcurve for a fading
  jet. {\em Lower:} This illustrates the possible combinations of
  on/off states for the fading jet and counterjet.  The squares
  coloured black are not formally possible, because of light-travel
  time effects while those in grey are formally possible but excluded
  for this object by observation.  The square labelled ``obs'' is the
  case we study in the text.
\label{fig:lightcurve}}
\end{center}
\end{figure}

What we actually observe is not just dependent on the kinematics of
light-travel time, but also on the evolution of the luminosity in the
jet material. To illustrate this point we consider the simple case
where the X-ray, ICCMB emission depends on the time since emission
from the central source in a manner qualitatively illustrated in
Fig.~\ref{fig:lightcurve}. 

In this picture the relic jet/counterjet plasma is moving at speed
$\beta_{\rm pl}$; this is very likely to be slower than the speed at
which the current jet or hotspots move otherwise we would not see any
relic plasma along the counterjet.  X-rays are emitted at a steady
rate for a period of time and then cool or fade over a timescale
$t_{\rm cool}$. When the cooling is sufficiently rapid, then the
cooling timescale $t_{\rm cool}<\Delta t$, and the leading edge of the
forward jet will be observed to cool and fade before the far extremity
of the counterjet is observed to cool. However, the details depend on
the rate at which X-ray ICCMB fades during the cooling epoch and in
the plasma rest frame.  Doppler boosting also plays a role, acting to
enhance the approaching emission relative to that which is
receding. The ratio of intensities for approaching and receding jets
is dependent on the flux density, $S_\nu\propto \nu^{-\alpha}$, with
$\alpha\approx 0.7$.  We consider the flux ratio between equal volumes
of plasma at either extremity of the approaching and receding jets,

\begin{equation}
{S_{\rm app}\over S_{\rm rec}} = \left({1+\beta_{\rm pl}\cos\theta}\over
{1-\beta_{\rm pl}\cos\theta}\right)^{3+\alpha}
{L_{\rm app}(t_{\rm app})\over L_{\rm rec}(t_{\rm rec})}
\end{equation}
where we have ignored factors $O(\beta_{\rm pl}^2)$.  

Initially, when neither jet has started to fade, the forward jet will
be more luminous on the basis of size relative to the receding jet and
Doppler boosting.  Subsequently, in the relic phase, both jets will
expand in their own rest frame and the Doppler boosting will be
modified according to the decaying, power-law spectrum and any
deceleration of the emitting material.  Then, as shown in
Fig.~\ref{fig:lightcurve}, the forward jet will be observed to fade
first and the flux ratio will be observed to decline, giving a much
more prominent counterjet. We measure the flux ratio for the relic
jet over the relic counterjet in the 2-10 keV band to be less than
0.25, as the upper limit to any jet emission is about 4 times smaller
than the measured counterjet emission.  From the total observed length
of the radio source of 130\,kpc, we calculate (assuming $\theta$ =
60$^{\circ}$) that the light travel-time difference between the
hotspots is 2 $\times$ 10$^5$ years.  Therefore the cooling time needs
to be $\lesssim$ 10$^5$ years.  Thus, provided that the intrinsic
luminosity of X-ray ICCMB declines quickly enough during cooling that
it takes place in a light-crossing time ($2 \times 10^5$ years), the
system will always evolve to a state where the receding jet has a
greater X-ray luminosity than the older, approaching counterpart. This
cooling timescale places a constraint on the particular cooling
mechanism for the jet which we consider in the next section.

\subsection{Cooling mechanism for the jet\label{sect:cooling}}

We require a mechanism that can produce a broad, X-ray counterjet with
a low energy turnover $\gamma$ $\lesssim\ 10^3$ needed to explain the ICCMB
emission.  The luminosity is required to decline in order to explain
the absence of an approaching X-ray jet in $2 \times 10^5$ years, by a
factor of four or more. With radio observations of an approaching radio jet as
motivation, we would expect the relic X-ray counterjet to have had a
thinner, radio jet as progenitor, containing a higher low-energy
turnover.  We consider two extreme cases for the possible mechanism,
first that the electrons in the radio jet escape diffusively into the
lobe and second that the jet expands adiabatically.

The diffusive escape of electrons from a radio jet of width $L$ would
occur over a timescale of order $L^2/D$ where $D$ is the diffusion
coefficient. Assuming that Bohm scattering is valid, we take the
particle mean free path to be equal to its gyroradius so that $t_{\rm
escape}\approx R^2/(r_{\rm g} c)$, where $R$ is the radius of the
relic jet width and $r_{\rm g}$ is the gyro-radius.  With $R\approx
7.5\times 10^{19}$m and $B\approx 10^{-9}$T such an escape time would
be many orders of magnitude greater than a Hubble time so that
diffusive escape could only deplete a radio jet of synchrotron
emitting electrons if their mean free path greatly exceeded the
gyroradius. A further problem with any model based on diffusive escape
is that it would provide no natural means of reducing the Lorentz
factors of energetic particles and/or magnetic fields, needed to
convert the plasma from a radio-synchrotron emitter to one dominated
by X-ray ICCMB.

An alternative model, which may work on sufficiently short timescales,
is that the radio jet simply expands adiabatically into the lobe.
This will reduce the particle energies and the magnetic field strength
leading to a decline in the radio synchrotron emission. The low-energy
turnover will also move to lower Lorentz factors, ultimately producing
sufficient $\gamma\sim 10^3$ particles needed for observable X-ray
ICCMB emission between $0.1$keV and $10$~keV.  Observationally the
X-ray counterjet has expanded in volume by roughly a factor of four
when compared with the younger radio jet. Under adiabatic expansion
$PV^{4/3}$ is constant, for material dominated by relativistic
particles; the factor of $4^{4/3}$ is plausible for the pressure ratio
between radio lobe and X-ray emitting material supporting the
adiabatic expansion model. While there is obviously scope for detailed
modelling of the X-ray counterjet, including the details of how the
expansion takes place and the possible excitation of shocks, it is
encouraging that the simple model of adiabatic jet expansion into the
lobe is consistent with the new observations presented in this paper.

The X-ray emission and upper limit to the radio flux allow us to
make estimates of both the number density of relativistic electrons
and the mean magnetic field.  In the X-ray counterjet we take the
differential electron spectrum, per unit volume, to be a power law
above a low-energy turnover $\gamma_{\rm min}$

\begin{equation}
N(\gamma) = N_0\displaystyle \left({\gamma\over \gamma_{\rm
    min}}\right)^{-p},\;\;\;\;\gamma\ge \gamma_{\rm min} 
\end{equation}
where we have normalised our spectrum such that $N_0$ is the
differential number of particles, per unit volume, at
$\gamma=\gamma_{\rm min}$. The observed X-ray counterjet luminosity
between photon energies of $0.1$keV and $10$keV is $L_X=1.4\times
10^{36}$\,W from a volume $V\approx 1.22\times 10^{61}\,{\rm m^3}$.
When the source of this emission is inverse Compton scattering, by the
power law distribution of electrons, of the Cosmic Microwave
Background, we can place a constraint on the
values of $N_0$ and $\gamma_{\rm min}$ (\cite{rybicki86}, section 7.3
and equation 7.31).

\begin{equation}
L_X\approx 1.4\times 10^{36}\,{\rm W}\,\nzero\gamthous^p
\end{equation}
where $p=2.4$. This value for $N_0$ determines the energy density,
$e$, in energetic electrons and the equipartition magnetic
field $(B_{\rm equip}^2/2\mu_0) = e$, giving

\begin{equation}
\bequipnt\approx 1.48\times 10^2\nzero^{1/2}\gamthous.
\end{equation}

However, the upper limit to the radio emission from the relic region
also places constraints on the magnetic field and the particle
spectrum. With a lower cut-off to the energetic electron distribution
there will be a minimum frequency to synchrotron radiation given by

\begin{equation}
\numin\approx 5\times 10^{-3}\gamthous^2\bnt.
\end{equation}

The upper limit to the flux density from the relic X-ray counterjet
regions is $P_\nu=2.048\times 10^{24}\,{\rm W\,Hz^{-1}}$ at a
frequency of $8$\,GHz, and this places a constraint on the field and
spectrum, determined by standard synchrotron radiation formulae
\citep{longair94} ,

\begin{equation}
\bnt^{(p+1)/2}\nzero\gamthous^p< 1.37\times 10^{-2}
\end{equation}

With the product $N_0\gamma_{\rm min}^p$ fixed by the observation of
X-ray ICCMB we find the important result that the actual mean B field
in the relic region is much smaller than that required for
equipartition

\begin{equation}
B\approx 10^{-4}B_{\rm equip}.
\end{equation}

Therefore, if the jet material was initially created with approximate
equipartition between field and particles, it is clear that the
magnetic field energy is dissipated more quickly than that of the
particles as the jet evolves into the relic phase. Clearly there are two 
important problems here; firstly whether, and how, the particles and field 
are produced in equipartition and secondly how their respective energy 
densities are dissipated. Detailed consideration of these issues are beyond the 
scope of this paper but are the subject of ongoing simulations for relativistic 
flows \citep[][]{reville07,spitkovsky08}. 

\subsection{Jet duty-cycle characteristics}

Taking the difference in light travel time to be $2 \times 10^5$
years, an (assumed constant) value of the hotspot advance speed of
0.1\,c and using equation\,2, gives a timescale of $10^6$ years
since the previous jet activity.

\subsection{Intermittency and relic jets in quasars and radio galaxies}

\cite{nipoti05} explored the intermittancy, hence duty cycle, of jet
activity (``flaring events'') in microquasars.  They suggested that if
there is a good analogy between microquasars and quasars, that
intermittancy should also be seen in the jet activity of quasars and
radio galaxies.  Indirect evidence the authors cited for this analogous
behaviour is the similarity of the duty cycle of microquasar jets with
the fraction of quasars that are radio-loud.  A specific test of this
prediction is that there should be evidence of previous/relic jet
activity in some quasars and radio galaxies.

The existence of relic jet-activity in Cygnus\,A, revealed via ICCMB
emission, indicates that there was an epoch of jet activity
earlier than the current synchrotron radio jet.  The
age of the current radio jet is of order $10^6$ years (Steenbrugge \&
Blundell, accepted companion paper).  There is no evidence of a bright
compact radio feature (e.g.\ resembling a hotspot) at the end of the
relic X-ray counterjet indicating there has been a pause between the
two jet episodes.  The fading timescale for synchrotron emission from
radio lobes is \lesssim\ $10^6$ years \citep{blundell00}, longer than
the light-travel time from the furthest extremity of Cygnus\,A to the
nearest.

Strong indications of episodic jet activity and re-starting jets have
been observed in the case of the so-called double-double radio
galaxies.  For example, \citet{schoenmakers00} derive an ``interruption
timescale'' of a few $10^6$ years between successive jet ejections.  

Why has a jet-cooling timescale not been established in other sources?
There are at least two reasons: (i) the duty cycle may be shorter in
other sources than in Cygnus\,A and (ii) the lack of suitably deep
observations.  We remark that the X-ray brightness of the relic X-ray
counterjet of Cygnus\,A, if it were redshifted to $z = 0.5$ and $z =
1$ respectively would be 0.16 and 0.079 of the observed luminosity at
redshift 0.057. To observe a relic counterjet at these higher
redshifts the exposure time would need to be increased by factors 9
and 18 compared to the current 200\,ks ACIS-I exposure time.

Nonetheless there are perhaps two other instances of a previous epoch
of jet activity in a powerful radio quasar being revealed by ICCMB,
rather than by lobe emission as in the case of the double-double radio
galaxies. The quasars are 3C\,294 where the current radio axis is
offset slightly from a faint X-ray axis (Fig. 4 of \cite{erlund06})
and is offset significantly from the bright X-ray axis (Fig.\ 2
\cite{erlund06}); another possible example is 3C\,356
\citep{crawford93}. 

\section{Conclusion}

We have analysed the X-ray counterjet revealed by the combined 200\,ks
{\it Chandra} ACIS-I and ACIS-S observations of Cygnus\,A. Its
power-law spectrum, with photon index of 1.7, indicates that the
feature is unlikely due to thermal gas, and is therefore
probably due to emission from jet plasma having spectral index
0.7. Comparing the X-ray detected counterjet with the observed radio
counterjet in the 5-GHz and 15-GHz radio images, we conclude that the
counterjet detected in X-rays is a relic jet. This conclusion was
reached from the following observations: (i) the curvature of the
outer parts of the X-ray counterjet is significantly different from
that of the current radio counterjet; (ii) this feature lacks any
directly associated radio emission implying a lack of high energy
synchrotron particles; and (iii) the width of the X-ray counterjet is
significantly broader than the radio jet or counterjet implying
expansion.

From the non-detection of an approaching X-ray (relic) jet, and the
light travel-time difference between the approaching and receding
hotspots, we find that the likely interval between the current and
previous episodes of jet activity is $\sim 10^6$ years and that the
cooling time of the jet has to be less than 2 $\times$ 10$^5$ years.
This short timescale can plausibly be explained by adiabatic expansion
causing the jet to cool and thereby fade in the X-rays. The upper
limit for observed radio emission due to the relic counterjet allows
us to deduce an upper limit for the magnetic field strength within
it; this is well below the equipartition value.  Our non-detections
of X-ray emission from the current jets, but the presence of X-ray
emission from a relic jet, indicates that there is a turnover in the
energy distribution of the jets ejected from the nucleus of
Cygnus~A. The jet emerging from the 
nucleus, is characterized by particles with Lorentz factors, $\gamma$
$\gtrsim$ 10$^3$.

\section*{Acknowledgments}
KCS would like to thank St John's College, Oxford for a fellowship;
KMB expresses her gratitude to the Royal Society and PD thanks the
Royal Irish Academy. 

\bibliography{references}

\label{lastpage}

\end{document}